\documentclass[preprintnumbers,amsmath,amssymb]{revtex4}

 \usepackage{remreset}

\usepackage[latin1]{inputenc}

\usepackage{graphicx}
\usepackage[bf]{caption2}

\newcommand{\ket}[1]{\mbox{\ensuremath{|#1\rangle}}}
\newcommand{\bra}[1]{\mbox{\ensuremath{\langle#1|}}}    

\newlength{\defbaselineskip}
\begin{document}

\title{Review on qudits production and their application to Quantum Communication and Studies on Local Realism}

\author{Marco Genovese}
 \email{m.genovese@inrim.it}

\author{Paolo Traina}%
 \email{p.traina@inrim.it}
\affiliation{%
I.N.RI.M -- Istituto Nazionale di Ricerca Metrologica\\
Strada delle cacce, 91 10135 Turin (Italy)
}%
\homepage{http://www.ien.it/ft/qm/index.shtml}
\date{\today}

\begin{abstract}
  The codification in higher dimensional Hilbert Spaces (whose logical
basis states are dubbed qudits in analogy with bidimensional qubits)
presents various advantages both for Quantum Information
applications and for studies on Foundations of Quantum Mechanics.

  Purpose of this review is to introduce qudits, to summarize
their application to Quantum Communication and researches on Local
Realism and, finally, to describe some recent experiment for
realizing them.

A Little more in details: after a short introduction, we will
consider the advantages of testing local realism with qudits,
discussing both the 3-4 dimensional case (both for maximally and
non-maximally entanglement) and then the extension to an arbitrary
dimension. Afterwards, we will discuss the theoretical results on
using qudits for quantum communication, epitomizing the outcomes on
a larger security in Quantum Key Distribution protocols (again
considering separately qutrits, ququats and generalization to
arbitrary dimension). Finally, we will present the experiments
performed up to now for producing quantum optical qudits and their
applications. In particular, we will mention schemes based on
interferometric set-ups, orbital angular momentum entanglement and
biphoton polarization. Finally, we will summarize what
hyperentanglement is and its applications.
\end{abstract}

\maketitle \vskip 0.5cm Keywords: {quantum communication, Bell
inequalities, QKD}

\section{Introduction}

Over the last decade, the possibility of manipulating single quantum
states as atoms or photons has largely increased opening, on the one
hand, new interesting possibilities in testing the Foundations of
Quantum Mechanics and, on the other hand, originating new disciplines
as  Quantum Information, which studies the encoding,
processing (Quantum Computation) and transmission (Quantum
Communication) of information using the properties of quantum
states, with promising advantages over classical systems.

In particular, Quantum Communication  has peculiar characteristics.
Purely quantum operations include, for instance, Quantum Key
Distribution (QKD), quantum teleportation, quantum dense coding and
entanglement swapping. These protocols stem from purely quantum
properties characteristics of Quantum Mechanics (QM), as the
uncertainty principle, the superposition principle, entanglement
(i.e., the existence of states whose wave function cannot be
factored in single particle wave functions) and the no-cloning
theorem. The possibility of the totally secure communication of a
cryptographic key (QKD) is of the utmost interest in the present
society, based on a widespread communications network. In fact, the
specific characteristics of QM allow a demonstration of absolute
security in communication based on quantum states. The basic
principle of QKD is sharing, by transmitting quantum states, a
random secret key between two distant parties. Various QKD protocols
have been proposed, each with its own advantages and disadvantages.
Most of them encode the information in a state which is defined in a
two dimensions Hilbert space (i.e., in a quantum bit, called qubit).
However, security can be increased by codifying in higher
dimensional Hilbert spaces (qudits).

Also in the researches on Foundations of Quantum Mechanics the use
of qudits presents advantages, in particular in relation to the
studies on local realism.

Incidentally, as we will discuss later, these two seemingly rather distant fields of research are tightly related to each other.

Purpose of this review is to introduce qudits (giving a wide
bibliography on them), to summarize their application to Quantum
Communication and Studies on Local Realism and, finally, to describe
some recent experiment for realizing them. A Little more in details:
In section II, we will consider the advantages of testing local
realism with qudits, discussing both the 3-4 dimensional case (both
for maximally and non-maximally entanglement) and then the extension
to an arbitrary dimension. In section III we will present the
theoretical results on using qudits for quantum communication,
epitomizing the outcomes on a larger security in Quantum Key
Distribution protocols (again considering separately qutrits,
ququats and generalization to arbitrary dimension). Finally, in
section IV we will introduce the experiments performed up to now for
producing quantum optical qudits and their applications. In
particular, we will mention schemes based on interferometric
set-ups, orbital angular momentum entanglement and biphoton
polarization; then, finally, we will summarize what
hyperentanglement is and its applications.

\section{Violations of Local Realism for qudits}

The quest for local realistic alternatives to QM has a
pluridecennial history, stemming from the 1935
Einstein-Podolsky-Rosen paper\cite{Einstein}, where completeness of
QM was questioned. Bell Inequalities (BI's) \cite{Bell,
clauserhorne, chsh} set a limit on the ability for any theory  based
on local realistic (LR) assumptions to reproduce all the statistical
results predicted by QM. Even if many experiments show violation of
BI's, no conclusive experimental test on local realism has been
performed yet, because those results rely on the accessory
assumption that the very few events observed, due to the finite (and
often low) efficiency of real detection apparatuses, are a faithful
statistical sample (this issue is often called {\em detection
loophole})\cite{GenoveseHVT, Santos, Garuccio, Pearle}. This means
that the possibility to find a Local Hidden Variable Theory (LHVT)
describing the quantum world (to which standard Quantum Mechanics
could be just an approximation) is not definitely discarded. At the
moment, only a  detection loophole free experiment has been
performed with Be ions \cite{beion}, but without fulfilling the
space separation requirement imposed in the original EPR
argument\cite{Einstein}. In order to close the detection loophole
one should have a detecting apparatus characterized by a minimum
detection efficiency $\eta^*$ whose value depends on the specific
measurement procedure and on the amount of local realism violation
predicted by Quantum Mechanics. As far as two-level quantum objects
are concerned, it has been estimated that a minimum quantum
efficiency $\eta^*=82.84\%$ is needed for a detection loophole free
test for maximally entangled qubits and that this value can be
lowered  to $\eta^*=66.7\%$ for non-maximally entangled qubits (see
\cite{eberhard,noi} and references therein).

In recent works it has been suggested\cite{massar1, massar2,
massar3}  that violation of local realism for {\em qudits} is
expected to grow with $d$ and this leads to a smaller required
$\eta^*$. In this sense it is interesting to explore Hilbert spaces
with dimension $d\geq2$. Incidentally, it is also worth to mention
that the quantification of violation of local realism is also
related to the quantification of entanglement as  a resource for
quantum information \cite{GenoveseHVT,manjpa}.

\subsection{Violation of Local realism for $d=3,4$}
The first papers  exploring the possible violations of local realism by qutrits (due to Kaszlikowski {\em et al.})\cite{qutritsvsqubits, chqutrits} were centered on the following scenario. 
Let us assume that the global state shared by two parties A and B
is a three-dimensional maximally entangled state as:

\begin{equation} \label{ququat}
\ket{\psi_{in}}=\frac{1}{\sqrt{3}}
(\ket{0}_A \ket{0}_B+\ket{1}_A \ket{1}_B+
\ket{2}_A \ket{2}_B).
\end{equation}

Now let 
both the parties send 
their particles 
through two similar devices called {\em unbiased 6-ports
BeamSplitters} (aka tritters) \cite{unbiasplitter}. 
These are a three-dimensional generalization of a 50\%/50\% Beam Splitter, whose 
 output state 
 in response to one of the three
inputs is a weighted superposition of its three eigenstates, so that
the global effect
, for instance, on particle A (B) defines
a unitary transformation with the followings entries:

\begin{equation} \label{multiport}
U_{kl}^A(\vec\phi^a)=\frac{1}{\sqrt{3}}e^{i(\frac{2\pi}{3}kl+\phi^a_l)}\quad
\quad(U_{kl}^B(\vec\theta^b)=\frac{1}{\sqrt{3}}e^{i(\frac{2\pi}{3}kl+\theta^b_l)})\quad k,l=0,1,2.
\end{equation}

In the last equation, the formal vector $\vec\phi^a$ ($\vec\theta^b$) refers to one
of two possible (identified by the value $a$ ($b$) of the upper index)
settings of three phase shifts used by A (B). For instance, $\vec\phi^1=(\phi^1_0,\phi^1_1,
\phi^1_2)$ ($\vec\theta^1=(\theta^1_0,\theta^1_1,
\theta^1_2)$).
The probability $P(A_a=k, B_b=l)$ of a given outcome for the two
quantities observed by A and B is calculated by projecting the
global output state on $\ket{k}_A\ket{l}_B$.

A measure of the strength of the violation of local realism was
defined as the minimal noise admixture $F_{max}$
\cite{resistance2noise} to the state below which no LHVT can
reproduce the correlation results. The general model of the mixed
qutrit state is
\begin{equation}
\rho_{noise}=(1-F)\ket{\psi}\bra{\psi}+\frac{F}{9}\textbf{1}
\end{equation}
where $0\leq F \leq 1$ and $\textbf{1}$ is the three-dimensional
identity matrix.

In a first paper \cite{qutritsvsqubits}, it is proven analitically that, for the described set of trichotomic observables with a specific choice of the phase shifts ($\vec{\phi^a},\vec{\theta^b}$), one has
\begin{equation}
F_{max}=\frac{11-6\sqrt{3}}{2},
\end{equation}
while for maximally entangled qubits this value is smaller
($\frac{2-\sqrt2}{2}$). Thus, entangled qutrits are more robust
against LR description than entangled qubits.

It must be pointed out that those results concern the computation of correlation functions.
In order to have a necessary and sufficient condition for the existence of local hidden variables, more strong constraints on outcome probabilities are needed, such as in CH inequality \cite{clauserhorne}. A general CH inequality \cite{chqutrits} for qutrits has been proposed as

\begin{equation}
\label{eq:CH3}
\begin{array}{l}
P(A_1=2;B_1=1)+P(A_1=2;B_2=1)-P(A_2=2;B_1=1)+P(A_2=2;B_2=1)+
\\P(A_1=1;B_1=2)+(A_1=1;B_2=2)-P(A_2=1;B_1=2)+P(A_2=1;B_2=2)+
\\P(A_1=2;B_1=2)+P(A_1=1;B_2=1)-P(A_2=2;B_1=2)+P(A_2=2;B_2=2)-
\\P(A_1=1)-P(A_2=2)-P(B_2=1)-P(B_2=2)\leq0.
\end{array}
\end{equation}
Numerical calculations on the outcome probabilities yield results that are equivalent to the previous ones for what concerns maximal noise admixture and consequently robustness against LR description.

Prompted by the results in qubits case\cite{eberhard}, computations
about the behaviour of more general non-maximally entangled qutrits
\cite{gisindurt2002, Genovesech} have been successively performed.
It has been shown that  non-maximal entangled qutrits exist which
violate local realism more strongly than maximal entangled ones (see
{\bf Fig. 1}
). It can also be shown that differences between the
violations of LR in the two cases increase with the dimension of the
system. It is also worth mentioning that it is always possible to
obtain any of these maximally ``non-LR'' states from the maximally
entangled one via {\em local operations} and {\em classical
communication} (LOCC).

{\bf Fig. 1}

Similar considerations can be extended to the $4$-dimensional
Hilbert Space \cite{mioviol} (several experiments related to the
measure of Bell Inequalities in three and four dimensions  have been
recently performed \cite{OAMqutrits,thew2004,fourph,avnyang} and
will be discussed later). The calculated values of $\eta^*$ and $F$
for $d=2,3,4$ are reported on table \ref{tab:confronto} and table
\ref{tab:rumore}, respectively.

{\bf Table 1}

{\bf Table 2}

\subsection{Violation of local realism for arbitrary dimension}
In one of the previously cited works \cite{gisindurt2002} it was
inferred that the violation of local realism increases with the
dimension of the system.  In order to investigate this tendency of
entangled qudit systems, a general class of Bell Inequalities  for
arbitrary dimension states have been proposed by Collins {\em et
al.} in 2002\cite{cglmp2002}. This $CGLMPI_d$ inequalities (whose
maximal values are limited by 2 for any LHV models) are obtained
starting from some logical constraints  that the correlation
function must satisfy for any local variable theory, following an
approach similar to the one leading to standard CHSH
inequality\cite{chsh}, to which, by the way, $CGLMPI_2$ is
equivalent. Also the values of the violations of such inequalities
by maximally entangled qudits are calculated for $d\geq2$

\begin{equation}
CGLMPI_d(\ket{\psi_{m.e.}})=4d\sum_{k=0}^{\frac{d}{2}-1}(1-\frac{2k}{d-1})(\frac{1}{2d^3\sin^2[\frac{\pi}{d}(k+\frac{1}{4})]}-\frac{1}{2d^3\sin^2[\frac{\pi}{d}(\frac{1}{4}-k-1)]})
\end{equation}
showing that the robustness of these states against LR description increases with $d$.

An asimptotic limit for $d\rightarrow\infty$ is also extrapolated ($CGLMPI_\infty$=2.296981).

A more detailed investigation on the structure of such inequalities is due to Chen {\em et al.} \cite{kwek2006} who,
in their $2005-2006$ works, calculated maximal violations and the corresponding eigenstates for $d$ up to $8000$.  The sets
of observables 
considered there are the extensions to any  dimension $d$ of the 
unitary transformations proposed by Kaszlikowski {\em et
al.}\cite{qutritsvsqubits, chqutrits} and the results are obtained
by maximizing the eigenvalues of the Bell operator. All the results
are in agreement with the previously cited papers and also, since it
is clearly evident that LR violations grow with $d$, it is
calculated that maximal violations are obtained once again for
non-maximally entangled qudits. Furthermore, an empirical formula
linking maximal violation value to $d$ and a family of convenient
approximate states are introduced.

\section{Qudits \& Quantum Communication}

Quantum Communication is the branch of Quantum Information Theory
dealing with the exchange of information among remote parties when
this information is encoded by means of some physical properties of
a quantum object ({\em e.g.}, single photon's polarization)
\cite{libro}. The most straightforward applications in this field
concern Quantum Cryptography\cite{gisin}.

In the most common scenario, a party, traditionally dubbed Alice
(A), wants to share some reserved information with a trusted
receiver, called Bob (B) and, in order to protect the security of
the communication against some malicious eavesdropper's (E) attacks,
they apply a {\em one-time pad} protocol (aka {\em Vernam Cypher}).
For this task they need, for each message, a system to share a
totally random key string (in principle of the same length of the
message). This is possible by means of Quantum Key Distribution
 protocols, with which A and B, only by virtue of the laws of
Quantum Mechanics, are able to produce perfectly correlated (and
unique) random strings to be used as encryption (or decryption) key
over messages exchanged in a public channel. Two main QKD classes
are known.

The first class stems from a protocol proposed by Bennet and
Brassard in 1984\cite{BB84} (BB$84$). In the original proposal, A
encodes her key bits by means of the polarization states of single
photons (for instance, vertical polarization corresponding to $0$
and horizontal one corresponding to $1$), randomly switching between
two {\em mutually unbiased} bases, such as $\{\ket{H},\ket{V}\}$
and $\{\ket{\frac{\pi}{4}},\ket{-\frac{\pi}{4}}\}$. B receives via
a Quantum Channel the qubits and he projects their polarization
states on either of the two bases, randomly as well. As a result,
when A and B use the same basis they get perfectly correlated bits,
while they get stochastic outcomes in the other cases. If E is
present, She can try to intercept A's photons, measure them and send
to B the eigenvector corresponding to the measured value. When this
strategy ({\em intecept-resend}) is used, it can be shown that E's
intervention modifies the probabilistic properties of B's string,
resulting in a increased Quantum Bit Error Rate (QBER) and a reduced
average mutual information shared by A and B ($I_{AB}$). More
sophisticated attacks can also be envisaged, for example E  could
entangle some probes to the transmitted qubits and perform a
measurement on them in a second time (coherent attack). In general
it can be proved that A and B can extract a secret key from a
corrupted one simply by classical algorithms (such as {\em error
correction} and {\em privacy amplification}) if the amount of their
mutual information $I_{AB}$ is bigger than E's one ($I_{EA}$,
$I_{EB}$).

The second class, related to a protocol first proposed by
Ekert \cite{Ekert91} (Ekert$91$), uses the correlations shown
 by entangled qubits as a guarantee of the confidentiality of the communication. In this case, A and B are separately given one of the two parties of a entangled bipartite system, and they randomly (and independently) project their qubits like in BB$84$. Since  E's presence spoils entanglement between A's and B's photons, whenever certain BI is not violated by randomly chosen subsets of their strings, communication is rejected, since it can be proven that the communication becomes unsecure in this condition.

Many experimental implementations of QKD protocols with qubits have
been performed. In this section we will briefly review the major
advantages stemming from the extension to arbitrary dimension
Hilbert spaces \cite{pasquiperes
,altri}.
\subsection{QKD with qutrits}
The immediate extension of BB84 protocols for qutrits has been proposed by Bechmann-Pasquinucci and Peres in $2000$\cite{pasquiperes}.
In a  $3$-dimensional Hilbert Space there are four possible {\em mutually unbiased bases} (MUOB's), corrisponding to 12 vectors. Assuming the first base is
\begin{equation}
\{\ket{\alpha},\ket{\beta},\ket{\gamma}\},
\end{equation}
the second is given by its discrete fourier transform
\begin{equation}
\{\frac{1}{\sqrt{3}}\left(\ket{\alpha}+\ket{\beta}+\ket{\gamma}\right),\frac{1}{\sqrt{3}}\left(\ket{\alpha}+e^{2\frac{\pi}{3} i}\ket{\beta}+e^{-2\frac{\pi}{3} i}\ket{\gamma}\right),\frac{1}{\sqrt{3}}\left(\ket{\alpha}+e^{-2\frac{\pi}{3} i}\ket{\beta}+e^{2\frac{\pi}{3} i}\ket{\gamma}\right)
\end{equation}
and the other two bases can be written as
\begin{equation}
\frac{1}{\sqrt{3}}\left(e^{\pm2\frac{\pi}{3} i}\ket{\alpha}+\ket{\beta}+\ket{\gamma}\right),
\end{equation}
and cyclic permutations.
A chooses randomly one of those vectors and sends it to B, who measures the state projecting on one of the four possible bases randomly as well. After that, B pubblicly reveals the basis used but not the outcome. If the bases are equal, A and B share a trit of information, otherwise they get fully uncorrelated results wich are immediately discarded. A and B perform as many iteration as they can in order to get a key of satisfactory length before proceeding with error correction and privacy amplification schemes (with sums {\em modulo} $3$ instead of parity checks).
In this case, considering a simple {\em intercept-resend} attack, E gets on the average $1/4$ of a trit of information for every transmission, while B error rate grows to $1/2$ (compared, respectively, to $1/2$ bit and to $1/4$ in the qubit case).
This protocol, compared with more complex ones, appears to be optimal for this kind of attack in the case of ``one way'' communications, since it gives the the lowest value of mutual information gained by E for a given quantity of information shared beetween A and B and the authors infer it to be optimal also for more sophisticated eavesdropping strategies.

Another possibility is to exploit $3$-dimensional BI's to perform
Ekert$91$-like protocols as proposed in \cite{ekqutr}. The proposed
setup is the same as the one previously introduced to describe
Kaszlikowski's CH inequality for qutrits and A and B, after the
transmission and the public declaration of the sets of phase shift
used, perform the calculation of the quantity on left side of eq.
(\ref{eq:CH3}) on a subset of their results. If maximal violation of
local realism is not reached they assume that something (E or noise)
had disturbed the transmission, otherwise they extract the key from
the remaining data. The authors show that, for a symmetric
incoherent attack in which E controls the source of entangled
qutrits, the correlation in the results between A and B is reduced
of a factor depending on the initial state prepared by E and on the
ancilla state used. E must keep this factor under a certain
threshold value ($\left(6\sqrt3-9\right)/2$) in order  to be
unnoticed, otherwise her presence will spoil the BI violation. But
computations yield that, in the region in which local realism is
violated, E's average error rate is always bigger than B's one and
the mutual information $I_{AB}$ shared by A and B is always greater
than information leaked to E (which is a condition for secret key
extraction). As a closing remark it is pointed out that this kind of
protocol is also more robust to noise than standard ones (Ekert$91$
and BB$84$) tolerating a noise admixture of $33.7\%$ (against
$29.2\%$).

\subsection{QKD with ququats}
A BB$84$-like protocol with ququats can be performed\cite{pasqui2000} with A choosing randomly among eight states belonging to two mutually unbiased orthogonal bases of $H_{4}$. Also in this case, given the first basis $\{\ket{\alpha_i}\}$, the second one ($\{\ket{\beta_j}\}$) can be obtained as discrete fourier transform of the latter.  Without E, if $n$ ququats are sent, on the average B guesses the right base in one half of the cases, so that $n/2$ shared ``quats'' ($n$ bits) are perfectly correlated. If E adopts a standard {\em intercept-resend} strategy, it is easy to show that on the average she gets half of the transmitted bits, just like in the analogous protocol in $2$ dimensions, but indeed E's presence induce a greater disturbance in the communication leading to an error rate in B's results larger than in the qubit case ($3/8$ against $1/4$).
Even if E uses an intermediate basis $\{\ket{\gamma_i}\}$ such that
\begin{equation}
\left|\left\langle\gamma_i|\alpha_i\right\rangle\right|=\left|\left\langle\gamma_i|\beta_i\right\rangle\right|=MAX\quad,\quad
\left|\left\langle\gamma_i|\alpha_j\right\rangle\right|=\left|\left\langle\gamma_i|\beta_j\right\rangle\right|=MIN,
\end{equation}
it can be shown that the average information leaked is less than a bit ($0.792$) for each
ququat transmitted and that the error rate introduced is even larger than in the previous case ($5/12$).

\subsection{Generalization to arbitrary dimension}
A suitable generalization\cite{cerf} of the BB$84$ protocol using
quantum $d$-level states, with $d$ an arbitrary integer, can be
realized following two ways. The first is to use only two mutually
unbiased orthogonal bases of $H_d$ (as in standard BB$84$), while
the second exploits all $d+1$ possible such bases (as in six-states
protocol). In both cases we suppose E owns some sort of cloning
machine (which must be considered imperfect due to no-cloning
theorem), used to copy the intercepted qudits before sending them to
B. Obviously this procedure adds a degree of mixedness to A and B
bipartite state. Let's firstly assume a protocol with only 2
complementary bases. If E strategy is based on individual attacks,
the best she can do (as in the qubit case) is to use a cloner that
copies equally (with equal fidelity $F_E$) both bases. Applying this
constraint to a general  class of cloning transformations
\cite{cloningtransf} and maximizing E's fidelity, it can be shown
that the information stolen by E equals the one shared by the
authorized parties when
\begin{equation}
F_E=F_B=\frac{1}{2}\left(1+\frac{1}{\sqrt{d}}\right)
\end{equation}
According to the known criterion this also gives the maximum noise
admixture $D=1-F$ below which the extraction of secret key via
privacy amplification is possible. This result suggests that
communication security grows with the dimension of the carrier
quantum state.

Following an analogous approach, it can be shown that in the case in which all $d+1$ MUOB's are used
(which are known only when $d$ is some power of a prime number) there is a slight improvement in the sense that
 for a given fidelity F, $I_{AB}$ is somewhat bigger and so is the maximum noise treshold D, thus the protocol is more robust than the previous one.

Also the case of coherent attack in which E performs a collective measurement on arbitrary long qudit strings is considered and it is found that  the upper bound to disturbance D in this case equals the value which is known for coherent attacks on strings of qubits\cite{coherent}.

Similar studies has been devoted to the investigation of possible extension of Ekert$91$ protocol to arbitrary dimension and its security against a wide class of attacks\cite{ekertqudits}. Also in this case an increased robustness against noise is found with respect to the standard qubit protocol.
\section{Implementations}

In this section we briefly review some major recent works on the
generation and exploitation of experimental entangled qutrits and
ququats for Quantum Communication purposes and LR tests.
Although the most of the reported schemes use single photons as
carriers, these applications rely on different physical
implementations corresponding to the different possible degrees of
freedom to be used in encoding the information.

\subsection{Postselected four-photon states}
In 2002, in order to give an experimental demonstration of the violation of CHSH inequality for spin-$1$ objects, polarization-entangled four-photon states produced by Type-II PDC\cite{Mandel} were studied\cite{fourph}. These states, which are present in the second order component of the downconverted field
\begin{equation}
\frac{1}{\sqrt{3}}(\ket{HH,VV}+\ket{HV,VH}+\ket{VV,HH}),
\end{equation}
are by all means four-photon qutrits and they can be observed by postselection techniques.

The polarization state was analysed with a setup similar to traditional Stern-Gerlach apparatus for spin-$1$ particles measurement and violation of CHSH inequality by more than $13$ standard deviation was observed.

\subsection{Interferometric schemes}
A first possible implementation of energy-time entangled qutrits for QKD is reported by Thew {\em et al.}\cite{thew2004}. In their scheme (an extension of Franson one\cite{Franson}
for qubits), they use the superposition state of the three possible paths of a single photon in a $3$-arm
interferometer as qutrit source.

Photons, obtained via Parametric Down-Conversion\cite{Mandel}
produced by a Periodically Poled Lithium Niobate waveguide, are
coupled to monomode optical fiber and then sent to an all-fiber
system which separates the photon pairs (by a beam splitter) and
addresses each of them to a different $6-port$ Michelson
interferometer, working as a tritter. The phase on each arm of those
tritters can be controlled via temperature. The path length
differences between the two interferometers are taken to be equal
and care is also taken to avoid polarization drift of photons inside
them. By six detectors (three for each interferometer) can be
observed nine possible combination of detection times distributions
corresponding to the possible stories of the produced photon pair.
These are reproduced in a histogram whose central peak  represents
the case in which both take the same path in each tritter (the
detection times coincide). Because the path length differences in the interferometers are
much smaller than coherence length of the pump beam
, no information is
available on creation times of photons or on the path taken, so
interference effects are detected (with visibility $V = 0.919 \pm
0.026$).

The authors claim to be able to produce high symmetry, maximally
entangled qutrits. Full characterization of the generated states is
performed and the declared fidelity is $F = 0.985 \pm 0.018$.

A similar scheme can be used to produce generalized GHZ states in three or four dimensions\cite{interferocina}.

Raising the dimension of this kind of scheme could be unpractical because it would require more than three
armed interferometers (anyway hyperentanglement could help).

An interesting scheme to produce high dimensional time-bin entangled
biphoton states has been proposed in 2005 \cite{fabriperot}. Here a
train of pulses generated by a mode-locked laser pumps a nonlinear
crystal. Since no information is available on the creation time of
photon pairs, the state created is of  the form
\begin{equation}
\ket{\psi}=\sum_{j=1}^N c_j e^{i\phi_j}\ket{j,j}
\end{equation}
where $j$ labels the time-bin, $c_j$ are constant probability
amplitudes, $\phi_j$ are the constant phase shifts between
successive pulses and $N$ is the number of pulses. The state is then
analyzed by a $2$-photon Fabri-Perot interferometer to demonstrate
high order entanglement. In this experiment the dimension is limited
only by the phase coherence between subsequent pulses and could, in
principle, be very large, even if it must be remarked that
experimental difficulties increase significantly as the
dimensionality grows.

\subsection{Orbital angular momentum entanglement}

In 2002\cite{OAMqutrits} Vaziri {\em et al.} have demonstrated for the first time
entanglement between qutrits using the Orbital Angular Momentum of
photons (which is conserved during PDC process \cite{pdcOAMcons}).
In this case the three possible eigenstates of OAM ($0$, $\hbar$,
$-\hbar$) span the qutrit space. After that a PDC pair is created,
each of the photons is addressed to an holographic module
(consisting of two displaced holograms) which can transfer the
incoming mode into a desired superposition $LG_{0l}$
LaGuerre-Gaussian Modes (each carrying $l\hbar$ per photon). After
the state is generated, each mode is sent to a device which can
discriminate among $LG_{0l}$ modes.

Using this scheme, a violation of more than $18$ standard deviations
of CGLMP inequality for three dimensions has been measured,
corresponding to non-maximally OAM entangled state.

Two years later\cite{Langford} complete characterization of OAM
entangled qutrits has been provided by Quantum Tomography measuring
the amount of entanglement and the degree of mixture and it has been
shown how to exploit such states for Quantum Communication (Quantum
bit commitment) with high security.

 Finally, Ren {\em et al.} experimentally showed  that high-dimensional
OAM entanglement of a pair of photons can be survived after a photon-plasmon-photon
conversion, the information of the spatial mode being thus coherently transmitted by surface plasmons\cite{renplasmon}.   
\subsection{Biphoton 
qudits}

Since the experimental Quantum Information scene is dominated by the generation, control and detection of polarization entangled biphoton states, the most natural experimental ground to reliably expand this discipline in more than two dimensions could be the generation of polarization entangled qutrits and ququats. It should also be underlined that this kind of states are definitely easier to control than the previously described ones because only linear optical elements are needed. On the other hand,  to experimentally tune all the superposition amplitudes in order to generate any arbitrary qutrit could be a very complex task if it is to be done by modifying a tritter phase shifts set or producing custom made holograms.

A first scheme addressed to the production of arbitrary polarization
qutrit states has been implemented in 2004 \cite{moscowqutrit}(see {\bf Fig. 2}) by
pumping with a single coherent source three non-linear crystals,
essentially overlapping Type-I and Type-II PDC emissions.  The
system allowed the generation of arbitrary qutrits and $12$ states
belonging to three mutually unbiased orthogonal bases of $H_3$ were
produced and measured with high fidelity (between 0.9842 and
0.9991). Anyway, this method does not allow to have entangled
qutrits.

{\bf Fig. 2}

A method\cite{mataloniq3ts} to engineer arbitrary pure and mixed
states of polarization qutrits and ququats exists for which only a
single nonlinear crystal and linear optical elements are needed. In
fact it is proven by means of {\em Singular value decomposition}
theorem that any arbitrary qutrit or ququat can be obtained
performing only unitary local transformations on a qubit bipartite
{\em seed} state (in this case a non-maximally entangled
polarization biphoton state).  Following this line it has been
possible to produce and measure with high average fidelity (0.93)
four MUOB's in $H_3$ and, in principle, five MUOB's in $H_4$ (20
states) can be observed as well.

Furthermore, an experimental way to produce and measure polarization ququarts based on non-degenerate biphoton field for QKD purposes has been lately  suggested\cite{moscowq4ts}.

In 2005 Neves {\em et. al.} showed to be able to achieve entangled biphoton qudits ($d=4$,$8$) also by means of the transverse spatial correlation between SPDC photons \cite{padua}. According to their scheme, Down-Converted  photons are sent to two identical apertures with $d$ slits and the $d$ possible paths (one for every slit) taken by each of the photons define the qudit space. For a proper adjustment of the pump beam, it is possible to have the two photons passing only by simmetrically opposite slits so that entanglement between the transverse spatial modes is obtained. The state has been generated and measured with high fidelity and it is proven to be a pure entangled state.

Finally, Ren {\em et. al} showed that the Hermite-Gaussian (HG) modes of SPDC downconverted beams are quasi-conserved and that the generated multidimensional biphoton states are HG modes entangled for some special cases \cite{renhg}.  

\subsection{Hyperentanglement}

A Hyperentangled state is a quantum system which is entangled
simultaneously in more than one of its degrees of freedoom.
Experimental generation of such states is interesting not only
because they can be seen as qudits to be used for tests on violations of
local realism \footnote{For the use of hyperentanglement in tests on the violation of local realism see also \cite{adan}.} or
QKD purposes, but also since Hyperentanglement, as it has been
suggested\cite{hyperBellmeas} and successively experimentally
proven\cite{mataloniBellmeas}, can provide a control qubit for for
complete Bell state discrimination with linear optical elements.

As an extension to a simple yet elegant scheme proposed by Brendel {\em et al.} in
1999\cite{brendel99} for producing energy-time entanglement, a source of energy-time and polarization Hyperentangled biphoton
states is proposed by Genovese and Novero\cite{Genoveseqc} (see {\bf Fig. 3}).

A single photon pulse feeds a Mach-Zender interferometer  in order
to generate, if the pulse duration is much smaller than the path
length difference between the {\em short} and the {\em long} arm, a
superposition state which reads:

\begin{equation}
\alpha\ket{short}+\beta\ket{long},
\end{equation}

where the complex amplitudes  of the superposition $\alpha$ and
$\beta$ are determined respectively by the coupling ratios at the
beam splitters and the phase difference introduced between the two
arms. If the previous photon pumps a non-linear crystal,
polarization entanglement can also be generated via PDC, resulting in
a global output biphoton state which can be written (for instance in
the type-II case) as

\begin{equation}
(\alpha\ket{short}+\beta\ket{long})\otimes\frac{1}{\sqrt2}(\gamma\ket{HV}+\delta\ket{VH}),
\end{equation}
where $V$ and $H$ label respectively vertical and horizontal polarization.

In order to observe interference in the time domain, one should
compensate for timing information introduced by the pulsed regime.
This is achieved by using other two identical interferometers as
detection apparatus, one for each PDC arm. Temporal
indistinguishability in this case is recovered by postselecting the
central detection peak out of the three possible, corresponding to
those events in which both downconverted photons of a pair travel
along the alternative arm with respect to the one taken by the
corresponding pump photon.

The proposed scheme  can  be used for implementations of QKD protocols
, entanglement swapping, teleportation and
generation of $3$-photon GHZ states.

{\bf Fig. 3}

Other ways\cite{interferocina} to combine known and efficient
entanglement sources to generate hyperentangled systems can be
thought: for instance, one can in principle exploit  simultaneously
energy-time (via interferometers) and OAM entanglement.

Also the previously discussed scheme by Mataloni {\em et al.} can be
extended by means of hyperentanglement\cite{mataloniBellmeas}. This
is done by a four-holes screen which allows to select two
complementary pairs of momentum modes from the incoming polarization
entangled biphoton field.

Finally, recently polarization-momentum hyperentanglment has been
used to perform experimental {\em all versus nothing} non-locality
tests for bipartite four-dimensional entangled
states\cite{avnyang,avnmataloni,val}, {\em i.e.} a demonstration of
non statistical predictions by quantum mechanics on the properties
of GHZ states.


\section{Concluding Remarks}

We have presented an introductory review on the codification of
quantum information in higher dimensional Hilbert spaces ($d>2$,
qudits) and application to quantum communication and studies of
local realism, both from a theoretical and experimental point of
view.

The results achieved up to now, and presented here, show that this
field of research is very promising for both these applications.

All the qudits realizations produced up to now are based on photonic
states. This is rather obvious when one is thinking to applications
to quantum communication and studies on local realism, where in
general quantum optical states play the game. Next steps will be on
the one side the development of efficient sources of photonic qudits
and on the other side the implementation of long distance quantum
communication channels where advantages of qudit respect to qubit
codification are exploited. Also an attempt to finally reach a
loophole free test of local realism could take advantage by the
reduced limit quantum efficiency required by using qudits.


In conclusion it must also be mentioned that recently new
applications of qudits, e.g. to quantum computation \cite{Qcomp} or
quantum communication protocols beyond QKD \cite{QC}, have been
proposed \footnote{Here one can also mention recent studies on
geometry of entanglement in qudits case, see \cite{g1} and
references therein.}. These results could prompt the implementation
of qudits on further physical systems as ions, atoms, etc. opening
new and unexplored sectors of research in this field.

\vskip 0.5cm
\textbf{Acknowledgements:} This work was supported in part by the
  by italian minister of research (PRIN 2005023443-002) and by
Regione Piemonte (E14).
\newpage

  \begin{figure}\label{chq}
   \begin{center}
   \begin{tabular}{c}
   \includegraphics*[scale=0.7]
   {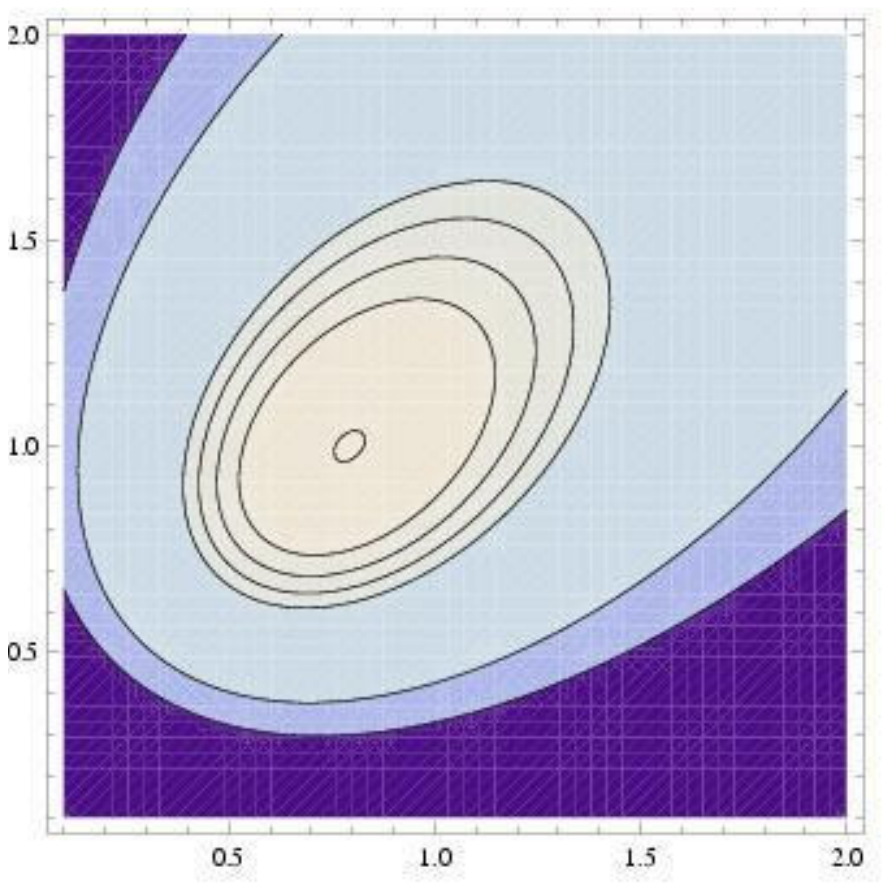}
   \end{tabular}
   \end{center}
   \caption{ \itshape{Contour plot of the left hand side of Eq. \ref{eq:CH3} calculated with the non-maximally entangled qutrit state
      $\Psi = {\frac{1}{\sqrt{1 + a^2 + b^2 }}} \left(| 0 \rangle _A |
0 \rangle _B + a | 1 \rangle _A | 1 \rangle _B + b | 2 \rangle _A |
2 \rangle _B  \right )$ in the plane a (abscissa) - b (ordinate).
Contour lines are at -0.15, -0.1, 0, 0.01, 0.02, 0.03, 0.05.}}
   \end{figure}

\newpage

\makeatletter
\renewcommand{\thetable}{\@arabic\c@table}
\makeatother
\begin{table}[htbp]
\textbf{Limit efficiency $\eta^*$ for $d=2,3,4$}\\
\centering
\begin{tabular}{||c||c|c|c||}
\hline
\textbf{$\eta^*(\%)$}&qubits&qutrits&ququats\\
\hline
\hline
CH&82.84 (66.7)&82.09 (81.39)&undefined\\
\hline
CGLMP&84.0896& 83.4357 (82.8330)&83.0993 (82.0237)\\
\hline
\end{tabular}
\caption{\itshape{Various expected limit quantum
efficiencies in order to close detection loophole
depending on qudits dimension (the bracketed
values concern non-maximal entanglement).}}
\label{tab:confronto}
\end{table}

\newpage

\begin{table}[htbp]
\textbf{Resistance to noise $F$ for $d=2,3,4$}\\
\centering
\begin{tabular}{||c||c|c|c||}
\hline
{}&qubits&qutrits&ququats\\
\hline
\hline
$F_{me}$&0.293&0.304&0.309\\
\hline
$F_{mv}$&0.293&0.314&0.327\\
\hline
\end{tabular}
\caption{\itshape{Resistance to noise for maximal ($F_{me}$) and
non-maximal ($F_{mv}$) entangled qudits depending on
dimension $d$.}}
\label{tab:rumore}
\end{table}

\newpage

\begin{figure}
\begin{center}
\begin{tabular}{c}
\includegraphics*[scale=0.8]{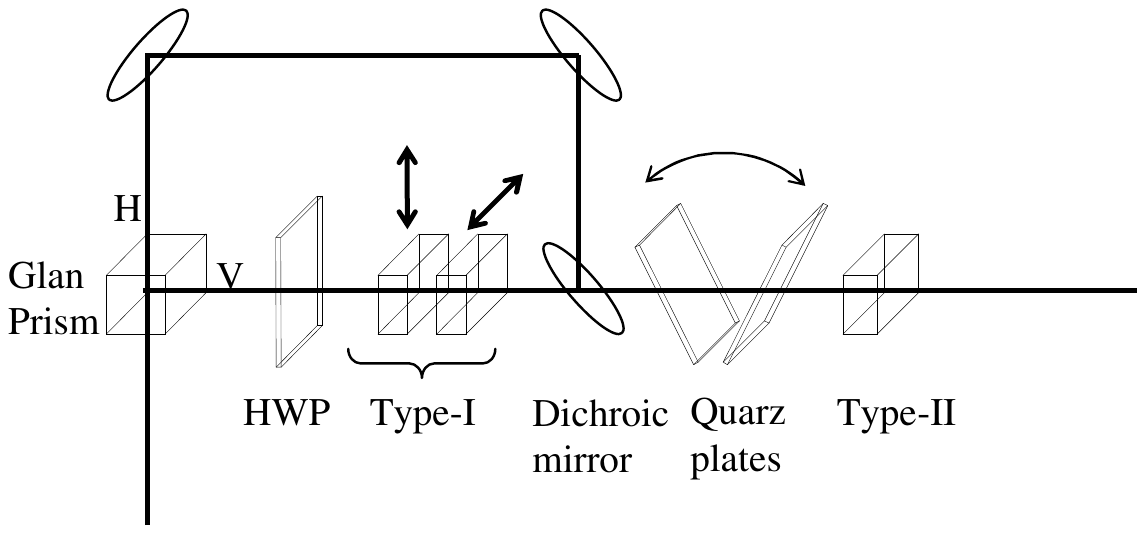}
\end{tabular}
\end{center}
\caption{\itshape{Scheme for the production of polarization biphoton qutrits. A Glan-tompson prism splits the vertical and horizontal components of the UV pump beam. The first component, rotated by a Half-Wave Plate (HWP), pumps two type-I crystals whose optic axes are orthogonal to each other, while the horizontal component pumps a type-II crystal. The output is the qutrit state given by the superposition of $\ket{VV}+ e^{i\phi}\ket{HH}$} (from type-I PDC) and $\ket{VH}$ (from Type-II).}
\end{figure}

\newpage

\begin{figure}
\label{figg:double}
\begin{center}
\begin{tabular}{c}
\includegraphics[scale=0.7]{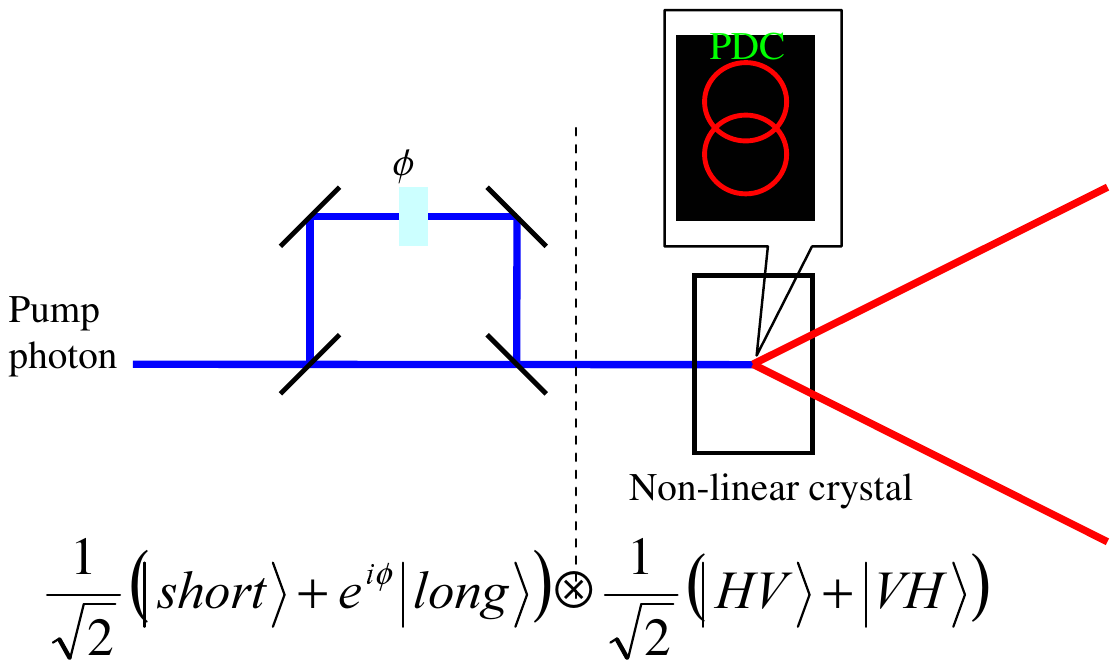}
\end{tabular}
\end{center}
\caption{\itshape{Scheme of a proposed source of ququats by means of hyperentangled 
entangled photon pairs: PDC process splits 
a single photon, in 
a superposition state corresponding to the two possible optical paths of 
the Mach-Zender, in a pair featuring both polarization and time-bin entanglement.
}}
\end{figure}


\end{document}